\documentclass[aps,
prl,twocolumn,showpacs,preprintnumbers,floatfix,
superscriptaddress]{revtex4}

\usepackage{graphicx}
\usepackage{dcolumn}
\usepackage{subfigure}
\usepackage{wrapfig}
\usepackage{cancel}
\usepackage{color}
\usepackage{textcomp}
\usepackage{amsmath}
\usepackage{amsfonts}
\usepackage{amssymb}
\usepackage{array}

\newcommand{\srbio}{SrBiO$_3$}
\newcommand{\babio}{BaBiO$_3$}
\newcommand{\tc}{$T_{\text c}$}

\newcommand{\ie}{\it{i.~e.}}

\newcommand{\etal}{\it{et~al.}}

\begin{document}

\title{Hybridization effects and bond-disproportionation in the bismuth perovskites}
\author{Kateryna Foyevtsova}
\affiliation{Department of Physics \& Astronomy, University of British Columbia,
Vancouver, British Columbia V6T 1Z1, Canada}
\affiliation{Quantum Matter Institute, University of British Columbia,
Vancouver, British Columbia V6T 1Z4, Canada}
\author{Arash Khazraie}
\affiliation{Department of Physics \& Astronomy, University of British Columbia,
Vancouver, British Columbia V6T 1Z1, Canada}
\affiliation{Quantum Matter Institute, University of British Columbia,
Vancouver, British Columbia V6T 1Z4, Canada}
\author{Ilya Elfimov}
\affiliation{Department of Physics \& Astronomy, University of British Columbia,
Vancouver, British Columbia V6T 1Z1, Canada}
\affiliation{Quantum Matter Institute, University of British Columbia,
Vancouver, British Columbia V6T 1Z4, Canada}
\author{George A. Sawatzky}
\affiliation{Department of Physics \& Astronomy, University of British Columbia,
Vancouver, British Columbia V6T 1Z1, Canada}
\affiliation{Quantum Matter Institute, University of British Columbia,
Vancouver, British Columbia V6T 1Z4, Canada}

\date{\today }
\pacs{74.20.Pq,74.70.-b,71.30.+h,71.45.Lr}

\begin{abstract}
We propose a microscopic description
of the bond-disproportionated
insulating state in the
bismuth perovskites $X$BiO$_3$ ($X$=Ba, Sr)
that recognizes the bismuth-oxygen hybridization
as a dominant energy scale.
It is demonstrated
using electronic structure
methods
that the breathing distortion
is accompanied
by spatial condensation of hole pairs
into local, molecular-like orbitals
of the $A_{1g}$ symmetry composed of O-$2p_{\sigma}$
and Bi-$6s$ atomic orbitals of collapsed BiO$_6$ octahedra.
Primary importance of oxygen $p$-states is thus
revealed, in contrast to a popular picture
of a purely ionic Bi$^{3+}$/Bi$^{5+}$ charge-disproportionation.
Octahedra tilting is shown to enhance the breathing
instability by means of a non-uniform band-narrowing.
We argue that formation of localized states upon breathing distortion
is, to a large extent, a property of the oxygen sublattice
and expect similar hybridization effects in other perovskites
involving formally high oxidation state cations.
\end{abstract}

\maketitle

The physics of perovskite compounds,
featuring $B$O$_6$ octahedra ($B$=cation) as building blocks,
is exceptionally rich.
The perovskites
can be driven through a variety of structural, electronic, and magnetic
phase transitions and
are hosts to such intriguing states of matter
as high-transition-temperature ({\tc}) superconductivity
[cuprates\cite{Bednorz86}, bismuth perovskites
$X$BiO$_3$ ($X$=Ba, Sr)
\cite{Cava88,Sleight75,Kazakov97}],
a pseudo-gap
state with strongly violated Fermi-liquid properties
(cuprates\cite{Timusk99}), and
a spin/charge density wave
[rare-earth nickelates $R$NiO$_3$ ($R$=rare-earth atom)
\cite{Medarde97}],
to name a few.
It is appealing to relate the diverse physical properties observed
across the perovskite family of materials with the individual
characteristics of the cation $B$. The latter can be magnetic (Cu
in the cuprates)
or not (Bi in the bismuth perovskites), orbitally active
(Mn in the manganites\cite{Tokura00})
or not (Cu, Bi),
prone to strong electronic correlations (transition-metal elements Cu and Ni)
or not (Bi).
With the due appreciation of the cation factor,
there are, however, many striking similarities
among different perovskite families suggesting
an equally important role of their common structural framework.
A vivid example is the transition
into a bond-disproportionated insulating phase found in both
the bismuthates
and the rare-earth
nickelates, in which oxygen plays an extremely important
role\cite{Mizokawa00,Park12,Lau13,Johnston14}.
Observations of this kind have given rise to
theories aimed at a unified description of the perovskites,
where the various competing phases emerge 
from polaronic and bipolaronic excitations of
the polarizable oxygen sublattice.

While it is a common practice within this approach
to assume that the only effect
of electron-phonon coupling is variation of on-site energies,
in this Letter
we demonstrate that
the effects due to hybridization between
the oxygen-$p$ orbitals and cation orbitals
can be even more important.
For this purpose, we focus on the bismuth perovskites $X$BiO$_3$
where the analysis is greatly facilitated
by the fact that the Bi ion valence states are non-magnetic,
relatively weakly
correlated, orbitally non-degenerate, and, to a first approximation,
not affected
by spin-orbit coupling (the Bi-$6p$ states
are above the Fermi level).
Yet, we believe that our findings
have broader implications for the perovskites in general.

The bismuth perovskites {\babio} and {\srbio}
are transition-metal-free high-{\tc} superconductors (when doped
with holes)
with intimately interlinked electronic and structural
phase transitions.
At low temperature, the parent compounds are
insulators with certain characteristic
distortions from an ideal cubic perovskite crystal structure.
Namely, the BiO$_6$ octahedra collapse
and expand alternately along all three cubic crystallographic
directions resulting
in {\it disproportionated Bi-O bond distances},
a so-called {\it breathing} distortion. Simultaneously, they rigidly rotate and tilt
to accommodate the Sr or Ba ions.
The insulating bond-disproportionated state
is often interpreted in terms of a charge-density wave.
It is speculated that the Bi ions, whose
nominal valency is $4+$, disproportionate
into Bi$^{3+}$ and Bi$^{5+}$\cite{Rice81,Cox76,Cox79,Varma88,Izumi07}
to avoid having a single electron in the $6s$ shell.
This picture, however,
is not supported experimentally\cite{deHair73,Orchard77,Wertheim82} and
is not consistent with the strongly covalent nature
of the Bi-O bonding on one hand and
the on-site repulsion effects on the other\cite{Harrison06,Mattheiss83}.
To stress the importance of these
two factors (covalency and on-site repulsion),
a useful analogy with the nickelates can be drawn.
There, the $3d^{7}$
configuration of Ni with one
electron in the $e_g$ doublet is similarly disfavored.
As was originally suggested by Mizokawa~{\etal}\cite{Mizokawa00}
and later shown in Refs.~\onlinecite{Park12,Lau13,Johnston14},
electronic correlations,
strong nickel-oxygen hybridization,
and a negative charge-transfer gap
result in holes preferring to occupy
oxygen sites rather than nickel sites.
In a combination with electron-phonon coupling
to a lattice distortion of the breathing type,
this promotes a ($d^8$\underline{L}$^2$)$_{S=0}$($d^8$)$_{S=1}$
configuration instead of the classical
charge-disproportionated ($d^6$)$_{S=0}$($d^8$)$_{S=1}$ state 
(\underline{L}=ligand hole, $S$=total on-site spin).

In the light of the above, here
we investigate the nature of
the insulating bond-disproportionated state
in the bismuthates from the perspective of strong Bi-$6s$/O-$2p$
hybridization
using density functional theory (DFT)
and local density approximation (LDA)\cite{SM}.
\begin{figure}[tb]
\begin{center}
\colorbox{white}
{\includegraphics[trim = 0mm 0mm 0mm 0mm, clip,width=\columnwidth]{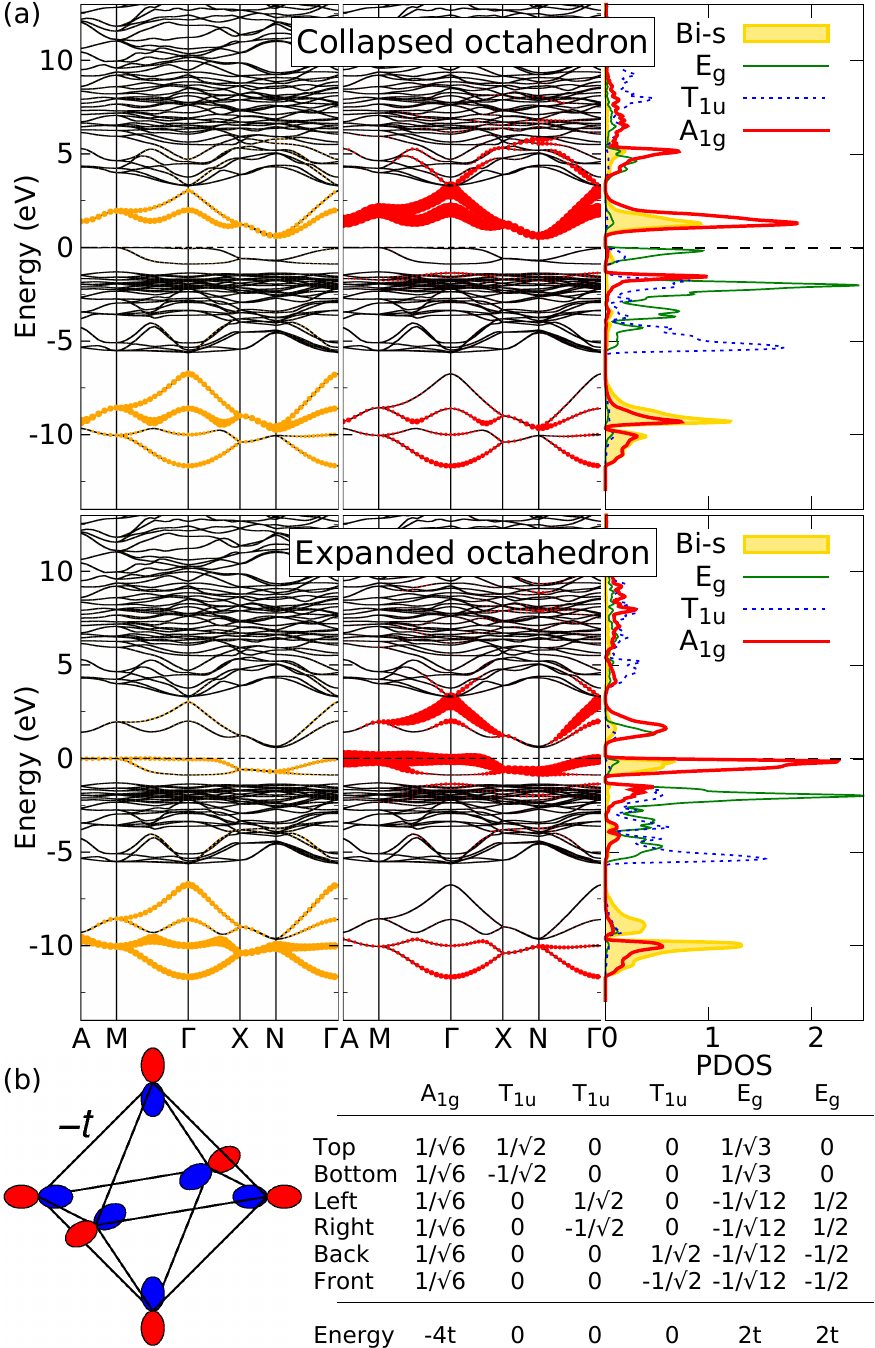}}
\end{center}
\caption{(a)~LDA electronic structure of {\srbio}
projected onto the Bi-$6s$ orbital and
combinations of the O-$p_{\sigma}$ orbitals
of a collapsed (top) and expanded (bottom)
BiO$_6$ octahedron. For the doublet $E_g$
and the triplet $T_{1u}$, only one
projection is shown. The Fermi level is set to zero,
and PDOS stands for projected density of states
and is given in states/eV/cell.
(b)~An octahedron of
O-$p_{\sigma}$ orbitals coupled via
nearest-neighbor hopping integrals $-t$ and
its eigenstates.
}
\label{F.bands2}
\end{figure}
\begin{figure}[tb]
\begin{center}
\colorbox{white}
{\includegraphics[trim = 0mm 0mm 0mm 0mm, clip,width=\columnwidth]{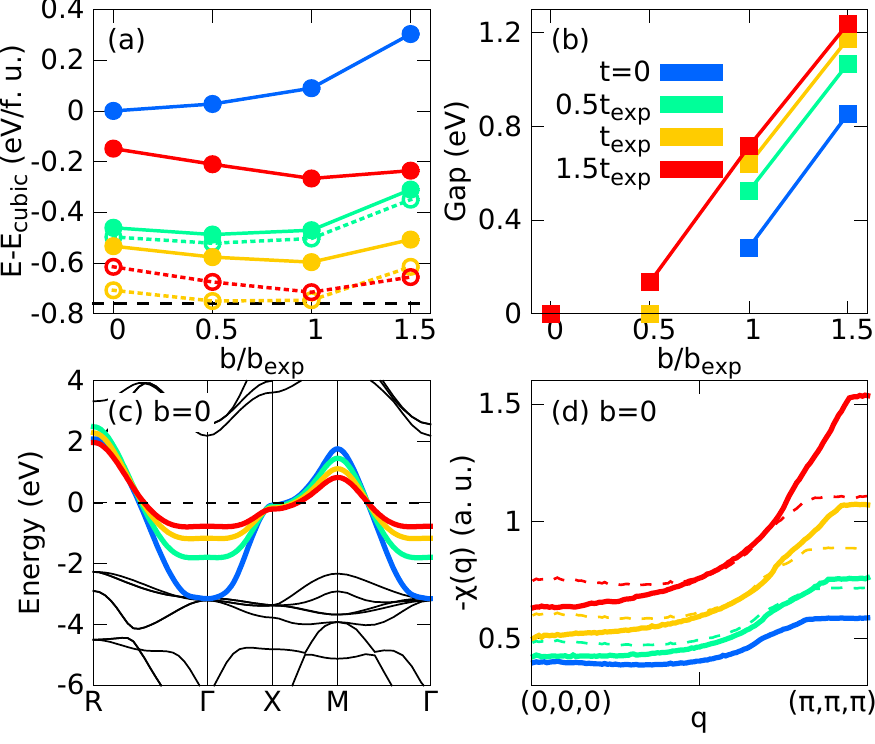}}
\end{center}
\caption{
(a), (b): LDA characterization of {\srbio} model structures
with varying degrees of the BiO$_6$ octahedra's tilting, $t$, and breathing, $b$:
(a) total energy per formula unit (f. u.)
and (b) charge gap.
In (a), solid lines and filled circles (dashed lines
and open circles) represent model structures
with fixed (relaxed) Sr atoms. The horizontal dashed line
marks the energy of the experimental {\srbio} structure.
(c), (d): The effect of tilting on (c) the half-filled band
and
on (d) the static susceptibility $\chi({\bf q},\omega=0)$,
at zero breathing.
In (d), solid (dashed) lines represent calculations
where non-linear effects due to tilting
are (are not) taken into account.
}
\label{F.str}
\end{figure}
\begin{figure*}[tb]
\begin{center}
\colorbox{white}
{\includegraphics[trim = 0mm 0mm 0mm 0mm, clip,width=2\columnwidth]{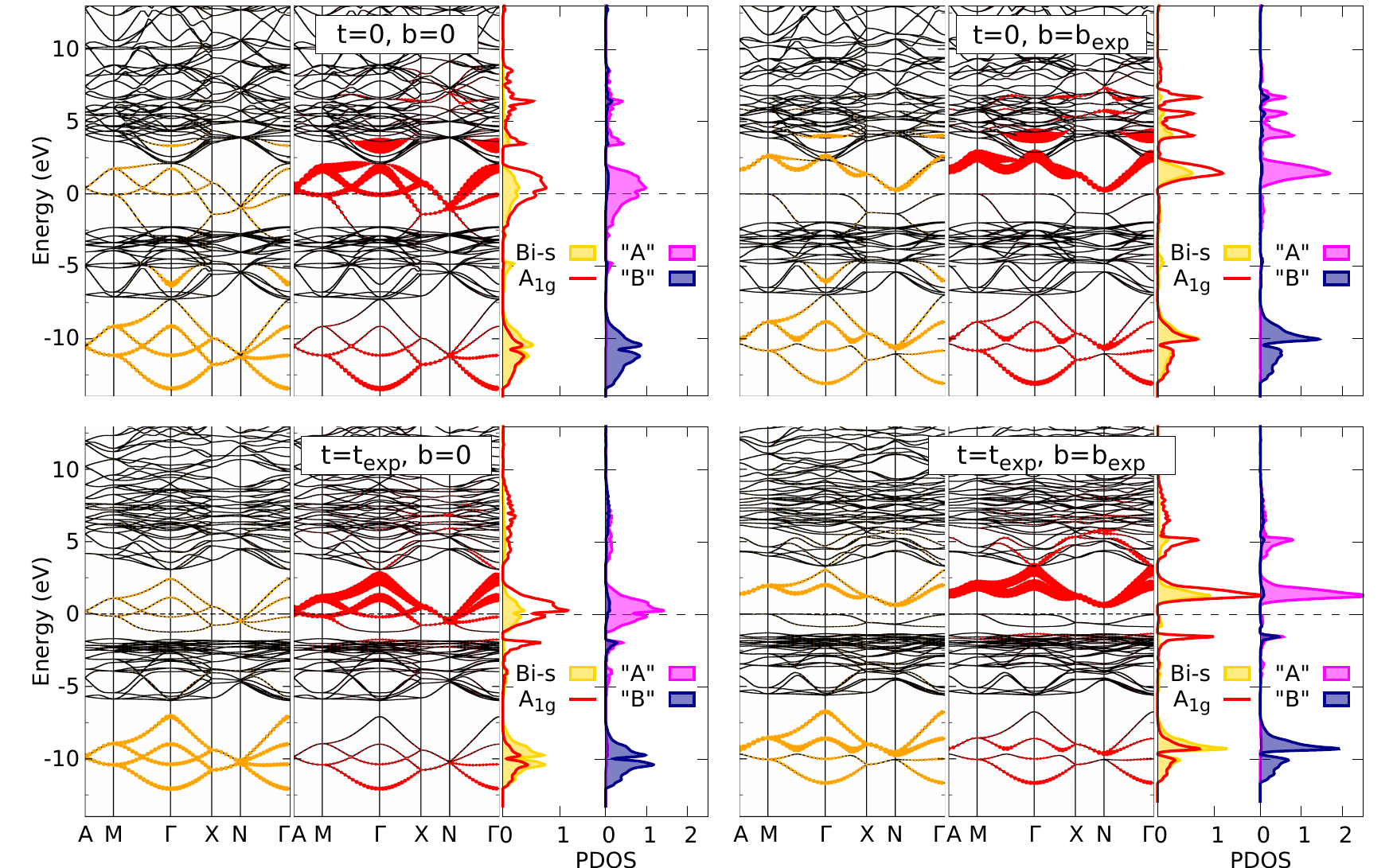}}
\end{center}
\caption{
LDA electronic structure of {\srbio}
as a function of breathing $b$ and tilting $t$.
Projections are made onto the Bi-$6s$ orbital and
the $A_{1g}$ combination of the O-$p_{\sigma}$ orbitals
of a collapsed BiO$_6$ octahedron,
as well as their bonding (``B'') and anti-bonding (``A'')
combinations.
}
\label{F.bands3}
\end{figure*}
We begin with highlighting basic features of the bismuthates' electronic
structure
\cite{Mattheiss83,Shirai90,Hamada89,
Liechtenstein91,Kunc91,Kunc94,Meregalli98,
Nourafkan12,Korotin12,Yin13,Korotin13}.
Since the LDA bandstructures of {\babio} and {\srbio}
are almost identical\cite{SM},
in the following
we will concentrate on
{\srbio}.
The Bi-$6s$
and O-$2p$ orbitals strongly hybridize to create
a broad band manifold near the Fermi level $E_F$ extending from
-14 to 3~eV [Fig.~\ref{F.bands2}~(a)].
The upper band of this manifold is
a mixture of the Bi-$6s$ orbitals and the $\sigma$ $2p$ orbitals of the oxygens,
{\ie}, the O-$2p$ orbitals with lobes pointing towards the central Bi\cite{comment}.
This upper band is half-filled as a result of {\it self-doping}:
the 18 $2p$ states of three oxygen ions and the two $6s$ states
of a Bi ion are short of one electron to be fully occupied.
Upon breathing, this band splits producing a charge gap.
We note that even in the bond-disproportionated
state, the bismuthates are very far from the ionic limit
of pure Bi$^{3+}$ and Bi$^{5+}$. The self-doped holes
reside predominantly on the oxygen-$p_{\sigma}$ orbitals,
a situation similar to that in the nickelates.
Below, we will infer the exact distribution of holes
by analysing the DFT data in more detail.

In our further analysis, we will be guided
by the following considerations.
First, cation valence states
of a given symmetry
will be strongly hybridized with only certain
combinations of surrounding O-$p$ orbitals, the Zhang-Rice singlet
in the cuprates
being an example\cite{Zhang88}. 
Thus,
a spherically symmetric Bi-$s$ orbital
should couple only to the $A_{1g}$  ({\ie},
fully symmetric) combination
of the $p_{\sigma}$ orbitals of the surrounding
oxygen atoms [Fig.~\ref{F.bands2}~(b)]. Hoppings to any other combination
of the O-$p$ orbitals cancel out by symmetry.
Second, since the Bi-$6s$ orbitals are
quite extended,
the hybridization between the Bi-$s$ and O-$A_{1g}$
orbitals is very strong and even becomes a dominant effect.
Third, in the bond-modulated phase, hybridization
within a collapsed (expanded) BiO$_6$ octahedron
is strongly enhanced (suppressed), resulting
in a formation of {\it local, molecular-like orbitals}
on the collapsed octahedra.

We are able to observe all of these effects in our DFT results
by projecting the LDA single-particle eigenstates onto
the combinations of O-$p_{\sigma}$ orbitals.
Let us first focus of the {\it collapsed}
octahedra. In the top panel of Figure~\ref{F.bands2}~(a),
one finds the Bi-$6s$ and O-$A_{1g}$ characters
at $\sim2$~eV and at $\sim-10$~eV. The difference, $\sim12$~eV,
is mainly a result of the hybridization splitting
between the bonding and anti-bonding
combinations, which indeed turns out to be the dominant energy scale
for {\srbio}.
The anti-bonding combination
is strongly peaked in the lowest two conductance
bands throughout the whole Brillouin zone.
In contrast, the location of the anti-bonding
$A_{1g}$ combinations of the {\it expanded} octahedra (bottom panel)
is rather diffuse, with some weight seen both below and above $E_F$
depending on the position in the Brillouin zone.
This indicates that the metal-insulator transition
with bond-disproportionation in the bismuthates
should be understood as a pairwise spatial condensation of
holes
into the anti-bonding
$A_{1g}$ molecular orbitals of the collapsed octahedra.
The small charge-disproportionation between the Bi ions
($\pm$0.15~$e$
inside the Bi muffin tin spheres)
appears to be a {\it marginal side effect} of such hole condensation.
Such a state,
which resembles the ($d^8$\underline{L}$^2$)$_{S=0}$ singlet
proposed for the nickelates,
was also hypothesized in Ref.~\onlinecite{Menushenkov01}.

It is interesting to trace the
evolution of the LDA projected density of states as a function of
breathing $b$ and tilting $t$.
For this,
we prepare a set of {\srbio}
model structures with varying
degrees of $b$ and $t$\cite{SM}.
In Figures~\ref{F.str}~(a) and (b),
the model structures are
characterized in terms of, respectively, the total energy
and the charge gap calculated within
LDA.
Parameters $b$ and $t$ are given with respect to
the experimentally observed distortions $b_{\text{exp}}$
and $t_{\text{exp}}$.
We find that a finite $t$
is required to stabilize the breathing
distortion.
Although this result is known to be
sensitive to details of DFT calculations\cite{Kunc91},
the absence of the breathing instability
at zero tilting can be qualitatively
understood in terms of a missing linear component
in the elastic energy of an oxygen atom positioned between
two Bi atoms.
With increasing $t$, the equilibrium $b$ shifts to higher values
[Fig.~\ref{F.str}~(a)],
while the charge gap opens sooner as a function of $b$ [Fig.~\ref{F.str}~(b)].
We conclude, in agreement with previous DFT
studies\cite{Liechtenstein91,Kunc91,Korotin12},
that tilting enhances the breathing instability,
through a mechanism yet to be discussed.
Interestingly, letting
Sr atoms relax from the high-symmetry positions (while keeping $b$ and $t$ fixed)
can significantly lower
the total energy for structures with high $t$
[open circles in Fig.~\ref{F.str}~(a)].
For the $(b_{\text{exp}},t_{\text{exp}})$ structure, for instance,
the energy drop is 0.15~eV per formula unit.

Figure~\ref{F.bands3} presents
projections onto collapsed octahedra
for the $(0,0)$, $(b_{\text{exp}},0)$,
$(0,t_{\text{exp}})$, and $(b_{\text{exp}},t_{\text{exp}})$ structures.
Here, the bonding (``B'') and anti-bonding (``A'') combinations
of the Bi-$6s$ and O-$A_{1g}$ orbitals are shown explicitly.
An important observation is that
even with no breathing a major portion of the anti-bonding
orbital weight is above $E_F$. Thus, the system is {\it predisposed}
to hole condensation. The breathing distortion just
makes collapsed octahedra more preferable for the holes to go to.
On the other hand, the primary role of the tilting distortion
is to reduce the bandwidth.
This band-narrowing, however, is very {\it non-uniform}
leading to certain important consequences,
as will be discussed later.
\begin{figure}[tb]
\begin{center}
\colorbox{white}
{\includegraphics[trim = 0mm 0mm 0mm 0mm, clip,width=\columnwidth]{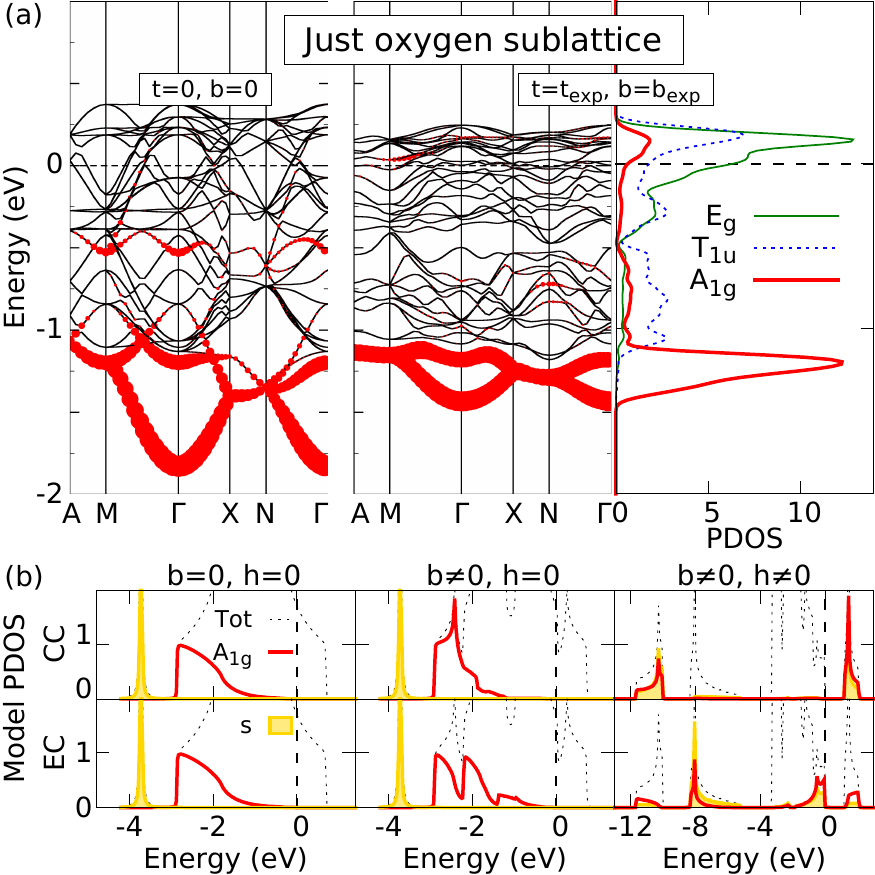}}
\end{center}
\caption{
(a) LDA electronic structure
of the oxygen sublattice of {\srbio}.
Projections are made onto
combinations of the O-$p_{\sigma}$ orbitals
of a collapsed O$_6$ octahedron.
(b) Model density of states
as a function of breathing $b$
and hybridization between $s$- and $p$-orbitals $h$.
CC (EC) stands for a collapsed (expanded) $p$-site cage.
The model states are 90\% filled,
{\ie}, there is one hole per $s$-orbital;
Fermi energy is set to zero
and marked with black dashed vertical lines.
}
\label{F.bands4}
\end{figure}

Formation of well-defined molecular O-$A_{1g}$ states on collapsed
octahedra is a property of the oxygen sublattice.
This is illustrated in Fig.~\ref{F.bands4}~(a)
which shows LDA calculations for an artificial
system consisting of only the oxygen sublattice of {\srbio}.
Here, however, the $A_{1g}$ states are
the lowest in energy and fully occupied.
It is the strong hybridization with the nearly degenerate Bi-$6s$ states
that pushes them above $E_F$.
Upon hybridization, the 2~eV broad
oxygen band and the essentially flat Bi-$6s$ band will
acquire a bandwidth of $\sim15$~eV. 
As a further illustration, we can consider a simple tight-binding (TB) model
for a two-dimensional perovskite-like lattice
of $p_{\sigma,\pi}$-orbital sites and $s$-orbital sites,
with nearest-neighbor $s-p$ and $p-p$ hoppings
as found in {\srbio}:
$|t_{sp\sigma}|\sim2.3$, $|t_{pp\sigma}|\sim0.3$, and
$|t_{pp\pi}|\sim0.1$~eV\cite{SM}.
Switching on the breathing
distortion helps to form molecular $A_{1g}$ states
associated with collapsed $p_{\sigma}$-orbital cages
at the bottom of the $p$-band
[compare the left and central panels of Fig.~\ref{F.bands4}~(b)].
Subsequent hybridization with the flat $s$ orbitals,
located right below the $A_{1g}$ states,
has a much stronger effect (since $|t_{sp}|\gg|t_{pp}|$) and
pushes the anti-bonding $s-A_{1g}$ combination
above $E_F$
[right panel of Fig.~\ref{F.bands4}~(b)].

Finally, we demonstrate the
significance of the non-uniform nature of the
tilting-induced band-narrowing.
For this purpose, we calculate
the static susceptibility $\chi({\bf q},\omega=0)$
with the help of a single-band TB model
for structures with zero breathing and varying tilting.
Two kinds of TB models,
describing the band crossing the Fermi level,
are considered:
(1) realistic $t\neq0$ models closely following the LDA bands\cite{SM}
and (2) $t\neq0$ models uniformly rescaled 
with respect to the $t=0$ case
such as to match the LDA bandwidth.
In Fig.~\ref{F.str}~(c), the band dispersions of the realistic models
are plotted in the Brillouin zone of a small cubic cell
so that the non-uniform nature of the $t$-induced band-narrowing
is particularly clear: the changes at $R$, for example, are much smaller than
at $\Gamma$. Using the realistic models and considering
all the energies spanned by the band, we
find a
dominating susceptibility
peak at ${\bf q}=(\pi,\pi,\pi)$ signaling breathing instability
that quickly grows with increasing tilting [Fig.~\ref{F.str}~(d), solid
lines].
This growth is much quicker
than it would be in the case of a uniform band-narrowing (compare
with the dashed lines). This indicates
that non-linear effects due to tilting,
such as, possibly, approaching perfect nesting conditions,
are very important for stabilizing the breathing distortion.

In summary, we have studied hybridization effects
in the bismuth perovskites and their
interplay with structural distortions such as breathing
and tilting. It is shown that strong hybridization between the Bi-$6s$
and O-$2p$ orbitals precludes purely ionic
charge disproportionation of a Bi$^{3+}$/Bi$^{5+}$ form.
Instead, the (self-doped) holes spatially condense into
molecular-orbital-like $A_{1g}$ combinations
of the Bi-$6s$ and O-$2p_{\sigma}$ orbitals
of {\it collapsed} BiO$_6$ octahedra,
with predominantly O-$2p_{\sigma}$ molecular orbital character. 
The
tilting distortion is found to strongly enhance
the breathing instability through (at least in part)
an electronic mechanism,
as manifested by a ${\bf q}=(\pi,\pi,\pi)$ peak
in the static susceptibility $\chi({\bf q},\omega=0)$.

It is expected that similar processes
take place in other perovskites as well
and can involve localized molecular orbitals of symmetries
other than $A_{1g}$. Thus, in the rare-earth
nickelates, the orbitals of relevance
would be the $E_g$ combinations of the O-$p_{\sigma}$ orbitals
hybridized with the Ni-$e_g$ orbitals.
Our model calculations
indicate that two-dimensional systems, such as cuprates,
can also exhibit hybridization effects of this kind.

\bibliographystyle{apsrev}


\begin{thebibliography}{35}
\bibitem[{com()}]{comment}
\bibinfo{note}{In Fig.~\ref{F.bands2}~(a), it appears as a set of four bands
  from -1 to 3~eV because of using a monoclinic
  $\sqrt{2}a\times\sqrt{2}a\times2a$ ($a$=cubic lattice constant) supercell of
  {\srbio} with four formula units.}
\end{thebibliography}

\end{document}